%% file: main.tex
\documentclass{ametsocV6.1}
\usepackage{gensymb}

\usepackage{color}

\title{Developing Machine Learning-Based Watch-to-Warning Severe Weather Guidance from the Warn-on-Forecast System}

\authors{Montgomery Flora\aff{2,3,4,*}\thanks{*Current Affiliation: The Weather Company, Atlanta, Georgia}, 
Samuel Varga\aff{1,2,3,**}\thanks{**Current Affiliation: Versik Analytics, Boston, Massachusetts}, 
Corey Potvin\aff{3,1,4},
Noah Lang\aff{4,5, ***}\thanks{***Current Affiliation: University of Wisconsin-Madison, Madison, Wisconsin}\correspondingauthor{Montgomery Flora, monte.flora@weather.com}}

\affiliation{\aff{1}{School of Meteorology, University of Oklahoma, Norman, Oklahoma} \\
\aff{2}{Cooperative Institute for Severe and High-Impact Weather Research and Operations, University of Oklahoma, Norman, Oklahoma} \\
\aff{3}{NOAA/OAR/National Severe Storms Laboratory, Norman, Oklahoma}\\
\aff{4}{AI Institute for Research on Trustworthy AI in Weather, Climate, and Coastal Oceanography}\\
\aff{5}{Valparaiso University, Valparaiso, Indiana}
}

\abstract{
While machine learning (ML) post-processing of convection-allowing model (CAM) output for severe weather hazards (large hail, damaging winds, and/or tornadoes) has shown promise for very short lead times (0--3 hours), its application to slightly longer forecast windows remains relatively underexplored. In this study, we develop and evaluate a grid-based ML framework to predict the probability of severe weather hazards over the next 2--6 hours using forecast output from the Warn-on-Forecast System (WoFS). Our dataset includes WoFS ensemble forecasts valid every 5 minutes out to 6 hours from 108 days during the 2019--2023 NOAA Hazardous Weather Testbed Spring Forecasting Experiments. We train ML models to generate probabilistic forecasts of severe weather akin to Storm Prediction Center outlooks (i.e., likelihood of a tornado, severe wind, or severe hail event within 36 km of each point). We compare a histogram gradient-boosted tree (HGBT) model and a deep learning U-Net approach against a carefully calibrated baseline generated from 2--5 km updraft helicity. Results indicate that the HGBT and U-Net outperform the baseline, particularly at higher probability thresholds. The HGBT achieves the best performance metrics, but predicted probabilities cap at 60\% while the U-net forecasts extend to 100\%. Similar to previous studies, the U-Net produces spatially smoother guidance than the tree-based method. These findings add to the growing evidence of the effectiveness of ML-based CAM post-processing for providing short-term severe weather guidance. 
}

\begin{document}

\maketitle

\statement{This study contributes to evidence supporting machine learning (ML) post-processing of storm-scale guidance to predict severe thunderstorms. We demonstrate that ML can be effectively leveraged to predict severe weather (i.e., tornadoes, large hail, and damaging winds) in the relatively unexplored 2--6 hour lead time range, highlighting its potential to fill the critical "watch-to-warning" gap in current forecasting capabilities.
}

\input{a-intro}
\input{b-methods}
\input{c-results}

\input{d-summary}


\clearpage
\acknowledgments
Funding was provided by NOAA/Office of Oceanic and Atmospheric Research under the NOAA-University of Oklahoma Cooperative Agreement $\#$NA2OAR4320204, U.S. Department of Commerce. This material is based upon work supported by the Joint Technology Transfer Initiative Program within the NOAA/OAR Weather Program Office under Award No. NA22OAR4590171. This material is also based upon work supported by the National Science Foundation under Grant No. RISE-2019758. The work began as an NSF-funded AI Institute for Research on Trustworthy AI in Weather, Climate, and Coastal Oceanography (AI2ES) Research Experiences for Undergraduates project (co-author Noah Lang) and led to a Master's project (co-author Samuel Varga). The authors thank Harold Brooks for informally reviewing an early version of the manuscript. The authors also acknowledge the team members responsible for generating the experimental WoFS output, which includes Joshua Martin, Kent Knopfmeier, Brian Matilla, Thomas Jones, Patrick Skinner, Brett Roberts, Nusrat Yussouf, and David Dowell. 

%
%
\datastatement
The training code is available at https://github.com/NOAA-National-Severe-Storms-Laboratory/frdd-wofs-ml-w2w. The HGBT guidance is run quasi-operationally and is available at https://cbwofs.nssl.noaa.gov/ under the Forecast tab. The data is not currently available in a public repository, but it is freely available upon request. 

\clearpage
\input{e-appendix}

\bibliographystyle{ametsocV6}
\bibliography{references}

\end{document}

%% file: a-intro.tex
\section{Introduction}

Forecasting severe weather like tornadoes, damaging winds, and large hail is challenging, especially in the watch-to-warning timeframe (the 0--6 hours preceding an event; \citealt{heinselman2024warn}). Convection-allowing models (CAMs) provide storm-scale guidance during this period but do not predict severe thunderstorm hazards directly. Forecasters use proxy fields like updraft helicity (UH) for tornadoes and simulated reflectivity for hail, requiring subjective interpretation and heuristic thresholds. While valuable, these methods lack probabilistic information, often target specific phenomena (e.g., tornadic supercells), and may not apply broadly to all severe weather-producing regimes. To improve short-term severe weather guidance, machine learning (ML) post-processing has emerged as a powerful tool for translating high-resolution CAM output into skillful probabilistic hazard forecasts \citep{Gagne+etal2017, Burke+etal2019, Sobash+etal2020, Loken+etal2020, Flora+etal2021, Loken+etal2022, Clark+Loken2022, McGovern+etal2023, UnetHailTobias}. 

The successful demonstrations of ML post-processing of CAM output for severe weather probabilities primarily focus on lead times of 0--3 hours or 12--36 hours. Lead times of 2--6 hours are largely unexplored, despite their importance for decision-making in the watch-to-warning continuum. This period poses a unique forecasting challenge: it exceeds the range of traditional nowcasting methods using radar and satellite trends, yet requires more precise spatial and temporal data than typical mesoscale or synoptic-scale guidance \citep{Brooks+etal1992_w2w}.

In this study, we use ML to post-process forecasts from the National Severe Storms Laboratory’s Warn-on-Forecast System (WoFS), which produces rapidly updating ensemble-based guidance at 0--6-h lead times. The WoFS has demonstrated skill in the watch-to-warning timeframe \citep{gallo2022exploring, heinselman2024warn}, but its raw ensemble output is uncalibrated and does not explicitly provide severe weather probabilities.  For earlier lead times (0--3 hrs), ML post-processing has already demonstrated success in the WoFS \citep{Flora+etal2021, Clark+Loken2022, UnetHailTobias}. However, given the limited predictability of thunderstorms \citep{Flora+etal2018}, it is unknown if and by how much ML can improve upon a carefully calibrated non-ML baseline at these lead times. 

To explore these questions, we train and compare traditional ML and deep learning (DL) models for predicting severe weather probabilities from 2--6-hr WoFS output. Previous research on post-processing CAM output has typically used traditional ML methods, especially random forests \citep{breiman2001random}, which capture non-linearities and are easy to tune \citep{Gagne+etal2017, Burke+etal2019, Sobash+etal2020, Loken+etal2020, Flora+etal2021, Loken+etal2022, Clark+Loken2022, hill2023new}. Here, we use gradient-boosted trees \citep{friedman2001greedy, LightGBM} (see Section~\ref{sec:ml_methods}\ref{sec:hgbt}), which are as skilled as random forests but significantly faster to train. Recently, \citep{UnetHailTobias} used a DL approach known as a "U-net" \citep{ronneberger2015u} (see Section~\ref{sec:ml_methods}\ref{sec:unet}) for predicting severe hail at 0--1 hr lead times using WoFS and observational inputs. Two advantages of U-Nets and other DL architectures include their automated feature extraction, which reduces manual feature engineering, and the speed at which they can be trained on modern GPUs. Given these advantages of DL methods and their growing adoption in meteorology, we develop and apply a U-Net model in this study in addition to the gradient-boosted trees model. 

To narrow the focus of the study, we train the ML and DL models to produce probabilistic forecasts of “any” severe weather akin to a Storm Prediction Center (SPC) Convective Outlook but at 2--6 hour lead times. A precursor to the present study explored individual hazard prediction \citep{VargaThesis}, but is omitted here for brevity. To assess the skill of the ML models, we develop a well-calibrated probabilistic baseline from the WoFS 2--5 km updraft helicity (UH) ensemble forecasts. We will show that both the traditional ML and DL models produce skillful guidance for severe weather at 2--6 hr lead times. 

%% file: b-methods.tex
\section{Data and Methods}

\subsection{Warn-on-Forecast Data}
The WoFS is a convection-allowing data assimilation (DA) and ensemble prediction system designed primarily for short-term thunderstorm forecasting. WoFS comprises 36 multi-physics analysis members, with the first 18 members used to initialize ensemble forecasts. The physics schemes used in the WoFS are detailed in \citet{skinner2018object}. WoFS assimilates radar and satellite data every 15 minutes to improve the accuracy of predictions of ongoing thunderstorms. Forecasts are generated at the top and bottom of the hour, with lead times of 6 and 3 hours, respectively. The forecast output frequency is every 5 minutes. Due to computational constraints, WoFS operates within a limited area of 900 x 900 km, which is repositioned for each case. As of 2023, WoFS can support multiple independent domains per case (non-overlapping). For the cases used herein, the WoFS DA was initialized using the High-Resolution Rapid Refresh DA output \citep{HRRR} at 1500 UTC, with the first forecasts typically initialized at 1700 UTC and running until 0300 UTC. Additional WoFS model and DA configuration information is available in \citet{heinselman2024warn}.

\begin{figure}[ht]
\centering
\includegraphics[width=0.9\textwidth]{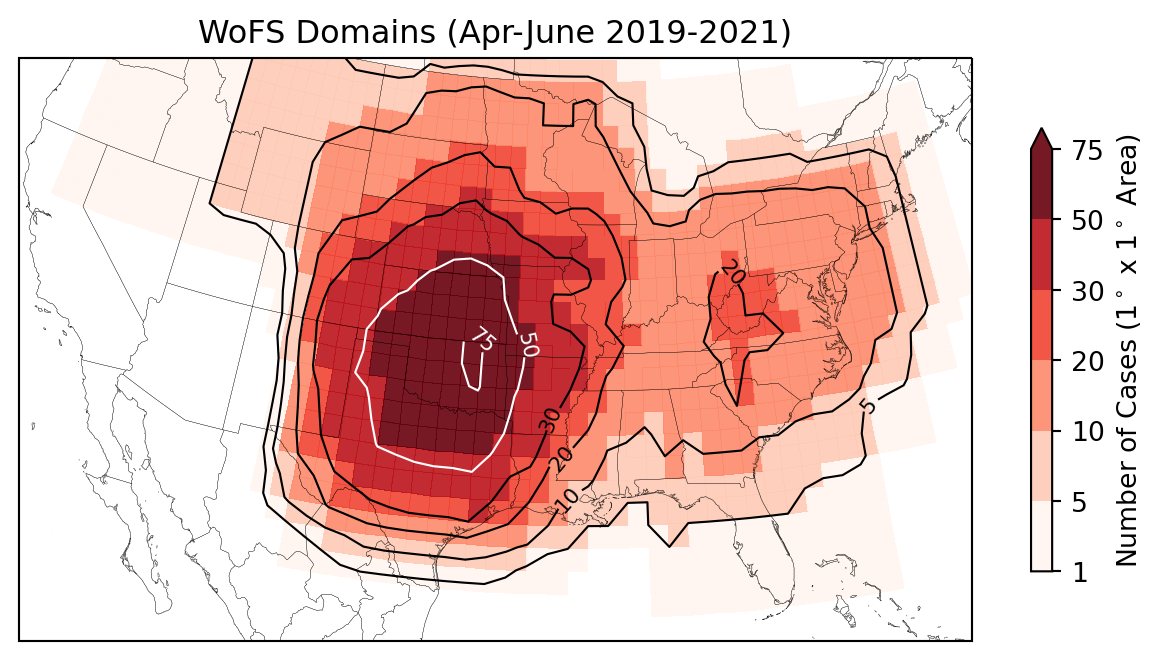}
\caption{ Heatmap showing the frequency of a 1$\degree$ × 1$\degree$ region within a 900 x 900 km WoFS domain during the 2019--2023 HWT-SFEs.}
\label{fig:domains}
\end{figure}

In this study, we utilize 108 cases generated during or just prior to the 2019--2023 Hazardous Weather Testbed Spring Forecasting Experiments (HWT-SFE) \citep{clark2023first} (Fig.~\ref{fig:domains}). Most cases sample the Central Great Plains, with a secondary maximum over the Mid-Atlantic (Fig.~\ref{fig:domains}). This sampling aligns with warm-season severe weather climatology \citep{SPC} and typical WoFS severe weather use cases. However, since samples are exclusively drawn from warm-season cases, the ML models trained on these data may not generalize well to year-round application. Since we target lead times of 2--6 hours, we are limited to forecasts initialized at the top of the hour (which extend to 6 h). There are 1121 forecasts in total. 

\subsection{Tabular and Deep Learning Data Processing Workflow}
\begin{figure}[ht]
\centering
\includegraphics[width=0.95\textwidth]{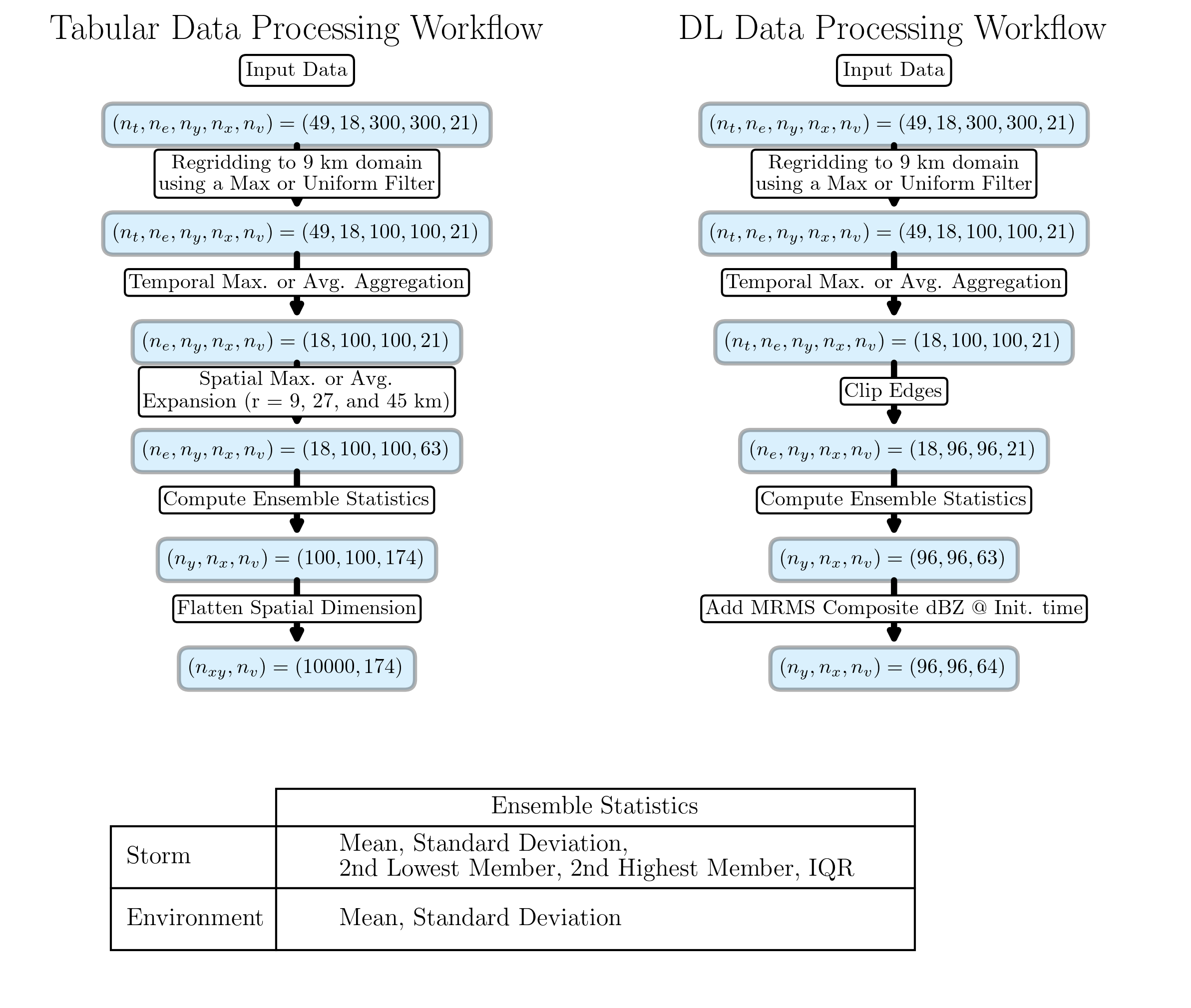}
\caption{ Illustration of the data pre-processing workflow for the tabular model inputs (left) and the U-net inputs (right). The data arrays sizes ($n$) for a single sample are given. The subscripts are $t$ = time, ($y,x$)= (latitude, longitude), $e$= ensemble, $v$ = variable/channel. The combined $xy$ indices indicate the common 2D array flattening to create tabular inputs. The ensemble statistics used are provided in the bottom table. }
\label{fig:data-workflow}
\end{figure}

The data preprocessing workflows for both the tabular and DL inputs are illustrated in Fig.~\ref{fig:data-workflow}. Table~\ref{table:input_vars} presents the 21 WoFS forecast fields used as inputs. For the DL inputs, we also incorporated the Multi-Radar, Multi-System (MRMS) composite reflectivity at model initialization time \citep{Smith+etal2016}, given the success of combining WoFS data with observation data for the U-net in \citet{UnetHailTobias}. However, we did not perform feature ablation to assess whether this addition contributed to the DL model's skill. We excluded the MRMS composite reflectivity from the tabular dataset, as it is unclear how best to condense this field into scalar statistics --- hence its inclusion in the DL model, which would ideally learn useful convolution kernels. 
\begin{table}[t]
\caption{Input variables from the WoFS. CAPE is convective available potential energy, CIN is convective inhibition, and LCL is the lifted condensation level.  HAILCAST refers to the maximum hail diameter from WRF-HAILCAST \citep{AdamsSelin+Ziegler2016, Adams-Selin+etal2019}. Parentheses indicate different vertical levels or layers where data were extracted. } 

\label{table:input_vars}

\begin{center}
\begin{tabular}{ll}
\hline \hline

Intra-storm & Environment   \\ 

\hline

Updraft Helicity (0--2 km, 2--5 km) & Lapse Rates (0--3 km, 700--500 mb) \\
Composite Reflectivity & 0--3-km Storm-Relative Helicity \\ 
0--2-km Average Vertical Vorticity & U Shear (0--6 km, 0--1 km, 3--6 km) \\
Column-maximum Updraft & V Shear (0--6 km, 0--1 km, 3--6 km) \\
80-m Wind Speed & Significant Tornado Parameter \\
HAILCAST & Supercell Composite Parameter \\
Okubo-Weiss Number & 75-mb Mixed-layer CAPE \\ 
                   & 75-mb Mixed-layer CIN \\

\hline
\end{tabular}
\end{center}
\end{table}

To be consistent with \citet{Flora+etal2021} and \citet{Clark+Loken2022}, tabular ML fields are classified as intrastorm or environmental. Intrastorm variables pertain to measures of a thunderstorm’s intensity, while environmental variables evaluate the broader, larger-scale dynamics and thermodynamics. Following Fig.~\ref{fig:data-workflow}, we initially coarsen the 3-km data from each ensemble member to a 9-km grid using a maximum (uniform) filter for the intrastorm (environmental) fields, followed by k-dimensional tree resampling using the nearest neighborhood method. The intrastorm fields are processed with a maximum filter as their distributions are highly skewed, whereas the environmental fields are smoother and closer to Gaussian distributions. Early in the project, we observed minimal skill loss when spatially coarsening the data, but it significantly improved computational speeds (not shown). For temporal aggregation, we compute the time maximum and average for the intrastorm and environmental fields, respectively. After spatial coarsening and time aggregation, the tabular and deep learning processing workflows diverge. We apply three spatial filters to the tabular data inputs to capture features at various spatial scales and implicitly measure ensemble spatial uncertainty. Intrastorm fields are processed using a spatial maximum filter, whereas environmental fields are treated with a spatially uniform filter. The radii of these filters are 9, 27, and 45 km, respectively. Ensemble statistics are computed from the spatially convolved fields; those statistics are available in the table of Fig.~\ref{fig:data-workflow}. Similarly, as mentioned above, more moments of the intrastorm variable ensemble distributions are computed than the environmental variables due to the skewness of the former.  All data are then flattened into a dataframe-like structure (rows and columns). For the deep learning workflow, we clip the domain edges from 100 x 100 to 96 x 96 after time aggregation. For the U-net architecture (see Section~\ref{sec:ml_methods}\ref{sec:unet}), it is beneficial for the input dimensions to be multiples of $2^{n}$ due to repeated applications of pooling operations that reduce the spatial dimensions by a factor of 2 at each layer. Then, the same ensemble statistics as for the traditional ML model are computed. We did not utilize different spatial filtering for the deep learning inputs because limiting the input feature size improves computational speeds, and since the U-net architecture is designed to capture signals at multiple spatial scales. After the data processing, the tabular dataset has 175 inputs, while the deep learning dataset has 63. 

Target fields are generated using National Centers for Environmental Information (NCEI) Storm Data. All reports of wind gusts (measured or estimated) exceeding 50 knots, hail size greater than 1 inch, and/or tornadoes are collected during the 2--6 hour forecast period and gridded to the WoFS domain. We also include reports within 15 minutes before or after the 2--6 hour forecast window to account for any timing inaccuracies in the reporting. The reports were expanded using a radius of 36 km, which aligns closely with a Storm Prediction Center-style (SPC) severe weather outlook. The resulting binary field indicates whether a grid point was within 36 km of the severe weather report. In earlier work, we found that smaller search radii of 9 and 18 km led to models with little or no skill (not shown).

\section{Traditional and Deep Learning Methods}\label{sec:ml_methods}

\subsection{Histogram Gradient Boosted Trees}\label{sec:hgbt}

Gradient-boosted trees (GBTs) are an ensemble method in machine learning \cite[][]{friedman2001greedy}. Training a GBT model begins with training a single decision tree to predict the target variables. The residual errors between this tree's predictions and the targets become the new targets for the next tree, and so forth. Thus, each tree learns from the residuals of the trees before it. During inference, the outputs from all trees are summed together. GBTs are generally as skillful as random forests for a wide range of problems but faster to train \cite[][]{bentejac2021comparative, LightGBM}. This study uses the HistGradientBoostingClassifier method available in scikit-learn \citep{pedregosa2011scikit}, which is similar to Microsoft's LightGBM \citep{LightGBM}. A histogram-based approach improves training speed by binning the input data into bins rather than evaluating each unique value for potential decision splits. Details on the hyperparameters used and the hyperparameter optimization are provided in the Appendix. 

\subsection{U-nets}\label{sec:unet}
\begin{figure}[t]
\centering
\includegraphics[width=1\textwidth]{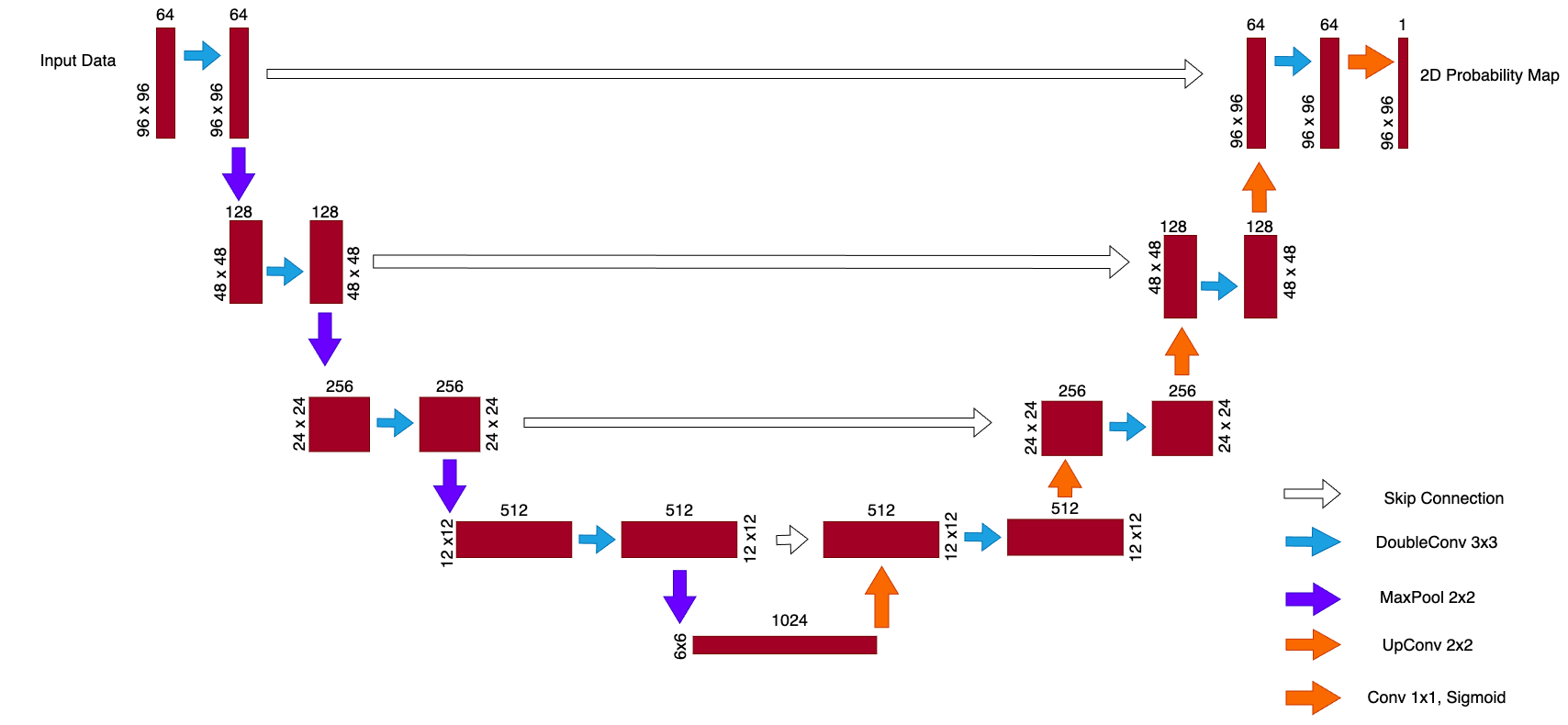}
\caption{Illustration of the U-net architecture used in this study. Image and channel sizes are provided on each rectangle's lower left-hand side and top. A DoubleConv layer comprises 2D convolution, batch normalization, ReLU, dropout, 2D convolution, batch normalization, and then ReLU again. For skip connections, the output of a given encoder layer is passed directly to the corresponding decoding layer and concatenated to the UpConv layer input. Skip connections are only connected across the U-net to the level with a similar spatial dimension.}
\label{fig:unet}
\end{figure}

U-nets, first introduced in \citet{ronneberger2015u}, are an image-to-image-based convolution neural network. The name is derived from the model's architecture, where input data is encoded through convolutions and pooling into a lower-resolution latent space and then transposed back into the original input shape (known as decoding; see Fig.~\ref{fig:unet}). The convolution filters used to process the data are learned during training. Given the vanishing gradient problem, output data from different layers are passed directly to the decoding layers, known as "skip" connections. These skip connections preserve fine-scale details by facilitating direct gradient flow, improving the network’s ability to reconstruct detailed outputs from compressed representations. This is a notably simpler approach than the skip connections used in the U-net++ and U-net 3+ architectures which connect to multiple layers \cite[][]{zhou2018unet++, huang2020unet}. The final convolutional layer uses one filter and a 1x1 kernel paired with a sigmoid activation function to output the positive class probability at each pixel. We implemented our U-net using PyTorch \citep{pytorch}. Details on the hyperparameters used and the hyperparameter optimization are provided in the Appendix.

\subsection{WoFS Baseline Method}
Following \citet{Flora+etal2021} and \citet{Clark+Loken2022}, we develop a rigorous non-ML baseline using WoFS neighborhood maximum ensemble probabilities (NMEP; \citealt{Schwartz+Sobash2017}). The ensemble probability is calculated as:
\[EP = \frac{1}{N}\sum_{n=1}^{N}f_i \geq t\]
where \(N\) is the number of ensemble members, \(f_i\) is the value of a forecast field, and \(t\) is some threshold. The EP represents the fraction of ensemble members with a forecast value exceeding the threshold. To obtain the NMEP, we apply a spatial maximum filter to each ensemble member before calculating the EP. For this study, we generate NMEPs from the 2--5-km updraft helicity (UH) field. UH has been frequently used as a baseline in other severe-weather-based ML studies (e.g., Gagne et al. 2017; Loken et al. 2020; Sobash et al. 2020; Clark and Loken 2022; Flora et al. 2021) since it is a skillful indicator of severe weather \cite[][]{sobash2016severe, sobash2020comparison}.

\begin{figure}[t]
\centering
\includegraphics[width=1\textwidth]{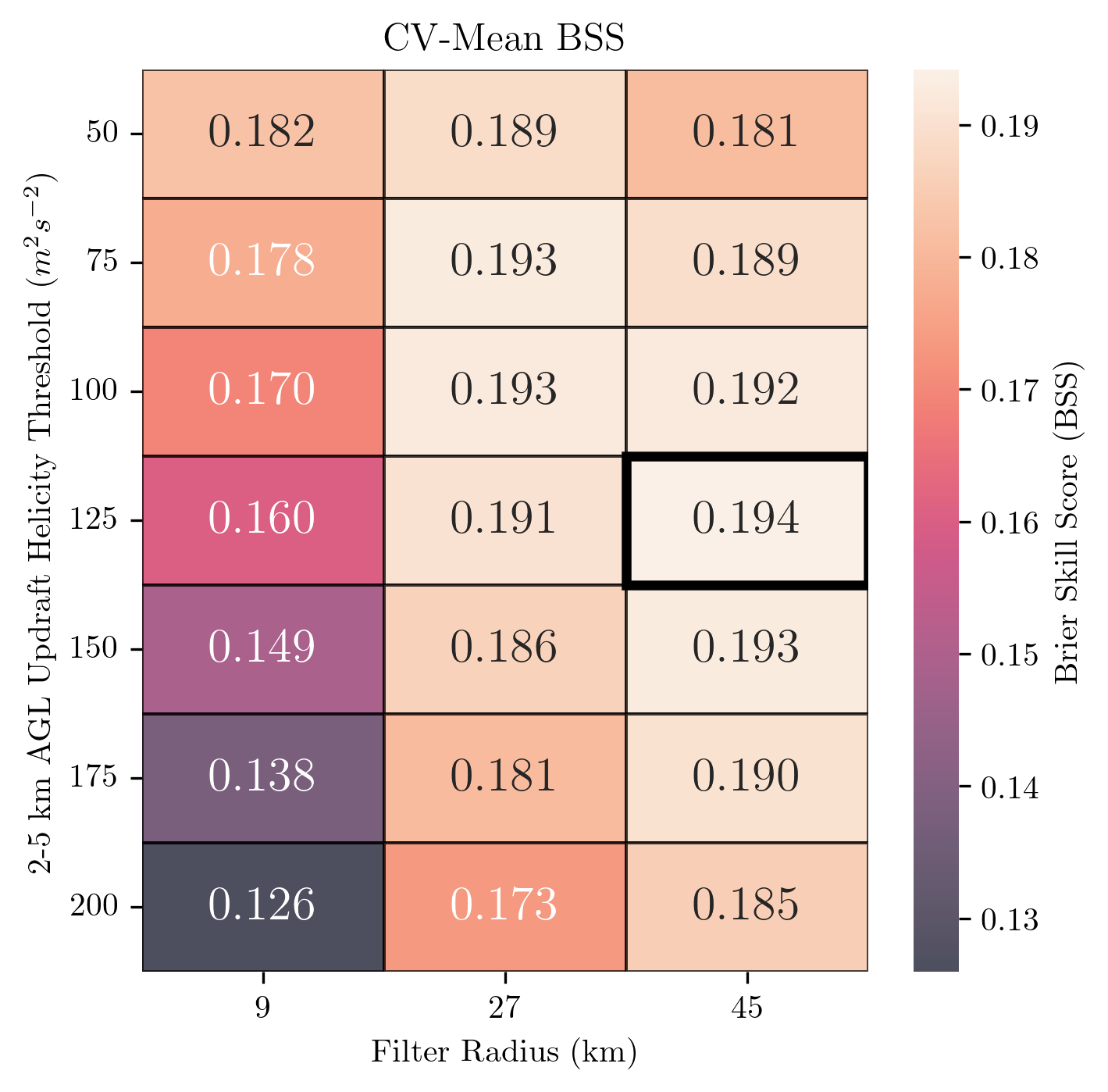}
\caption{Heatmaps of the NMEP cross-validation (cv)-mean Brier Skill Score (BSS) for choice of 2--5 km UH threshold and neighborhood size. The BSS is calculated as the mean BSS across all validation folds using 5-fold cross-validation on the training set. The highest BSS is highlighted with a black outline. 
}
\label{fig:heatmaps}
\end{figure}

The UH threshold and neighborhood size for the NMEPs were determined through 5-fold cross-validation (CV) on the training dataset, using the same approach as in \citet{Flora+etal2021}. Figure \ref{fig:heatmaps} shows the CV-mean Brier skill score (defined in Section~\ref{sec:ml_methods}\ref{sec:metrics}) for each threshold and neighborhood size. A threshold of 125 $m^2s^{-2}$ and a radius of 45 km produced the highest BSS (0.194). However, other combinations produce a similar BSS, so the BL performance is relatively insensitive to the choice of UH threshold or neighborhood radius. As in \citet{Flora+etal2021}, we use the isotonic regression model available in scikit-learn \citep{Pedregosa+etal2011} to improve the calibration of the baseline predictions. This is done using the approach of \cite{platt1999probabilistic}, where the validation data from each cross-validation fold is concatenated and used to learn the calibration function. 

\subsection{Model Evaluation Methods and Metrics Used}\label{sec:metrics}

For model development and evaluation, the full dataset was split into 75 cases for training and 33 cases for testing. A random date-based split is consistent with prior WoFS-based ML research \citep{Flora+etal2021, Clark+Loken2022, PotvinMLErrors2024}. The training sample size was 783 (in terms of patches; 7.21M grid points) for the deep learning model and 8.04M gridpoints for the tabular model. Regarding gridpoints, the training sample sizes differ as some cases are missing MRMS data, slightly reducing the DL input size. To ensure a fair comparison between the tabular and deep learning models, evaluation was limited to a 96 x 96 area, resulting in a testing sample size of 3.11 million points. Model development, which includes training, hyperparameter tuning, and calibration training, was carried out on the training dataset. All metrics described below were calculated on the held-out testing dataset and were bootstrapped ($N$=100) to generate confidence intervals for significance testing. Given that forecasts from successive initialization times are not entirely independent, we employed block bootstrapping. Initialization times were organized into four blocks, with times in each block separated by 4 hours. Data is randomly selected with replacement from one block during each bootstrap iteration. Permutation testing was used for statistical significance. Metrics and curves presented below are the bootstrap mean values. 

We used two well-known diagrams and their corresponding metrics to evaluate model performance: the performance diagram \citep{roebber2009} and the reliability diagram \citep{Hsu+Murphy1986}. The following provides a brief description of each diagram. 

Before describing the performance diagram, we define the components of the 2x2 contingency table. Forecast "yes" or "no" is defined as greater than or equal to and less than some probability threshold, respectively. Target "yes" or "no" is defined as whether severe weather occurred. The four components of the contingency table are 1) "hits" ($a$): forecast "yes" and target "yes", 2) "false alarm" ($b$): forecast "yes" and target "no",
3) "misses" ($c$): forecast "no" and target "yes", and 4) "correct negative" ($d$): forecast "no" and target "no". From these values, we can compute the probability of detection (POD = $\frac{a}{a+c}$), success ratio (SR=$\frac{a}{a+b}$), critical success index (CSI; $\frac{a}{a+b+c}$), and frequency bias ($\frac{a+b}{a+c}$).

The performance diagram, also known as the precision-recall diagram \citep{roebber2009, brooks2024rose}, plots POD and SR for a series of probability thresholds. The resulting curve measures the tradeoff between missing events and false alarms and illustrates how changing the decision threshold impacts their balance. Diagnosing the tradeoff between misses and false alarms is particularly important for imbalanced prediction problems (e.g., base rates $<$5-10\%), where misses are often more costly than false alarms. As a result, more emphasis is placed on correctly predicting events rather than nonevents \citep{Davis+Goadrich2006}. Similar to receiver operating characteristic curves (ROC) \citep{Metz1978}, the area under the curve is a useful metric to summarize the performance diagram curve. However, unlike the ROC area, there is no agreed-upon interpretation of this area. The area under the performance diagram curve (AUPDC; \citealt{Boyd+etal2012, Flora+etal2021, Miller+etal2021, flora2024exploring, Flora+etal2025}) is computed as the weighted average of SR \citep{Boyd+etal2012}:
\begin{equation}
    \text{AUPDC} = \sum_{k=1}^K (\text{POD}_k - \text{POD}_{k-1})\text{SR}_k
\end{equation}\label{eqn:aupdc}
where POD and SR are the probability of detection and success ratio, and $K$ is the number of probability thresholds used to calculate POD and SR. AUPDC is computed using the average precision score in scikit-learn \citep{Pedregosa+etal2011}, which utilizes each unique predicted probability for $K$ in equation 1. Given the sensitivity of AUPDC to base rate, we use the normalized AUPDC (NAUPDC; \citealt{Flora+etal2021}):
\begin{equation}
    \text{NAUPDC} = \frac{\text{AUPDC} - \text{AUPDC}_{min}}{1 - \text{AUPDC}_{min}}
\end{equation}
where $\text{AUPDC}_{min}$ = $c$ is the base rate\footnote{Base rate is defined as the percentage of positive class samples in the dataset.} of the dataset. 

Another standard metric associated with the performance diagram is the critical success index (CSI; \citet{brooks2024rose}). For this study, we compute the maximum CSI and normalize for base rate using an equation similar to (2), yielding the normalized CSI (NCSI). 

The reliability diagram assesses if the predicted probabilities are calibrated, as predictions must reflect observed frequencies to be considered probabilities. We bin the predictions to create the reliability curve, then use the corresponding target value pairs to determine observed frequencies for each bin. A model is perfectly calibrated if the average prediction in each bin aligns with the observed frequency in that bin. The 1-to-1 line is often plotted to indicate this perfect calibration. The key metric associated with the reliability diagram is the Brier score (BS), which calculates the mean squared error between predicted probabilities and the binary target value. The Brier score can be converted into a skill score (Brier Skill Score, or BSS) by comparing it to a score obtained from a low-skill baseline, typically a constant climatological prediction. This study uses the testing dataset base rate as the climatological prediction. 

%% file: c-results.tex
\section{Results}

\subsection{Example Forecasts}
\begin{figure}[h]
 \centerline{\includegraphics[width=35pc]{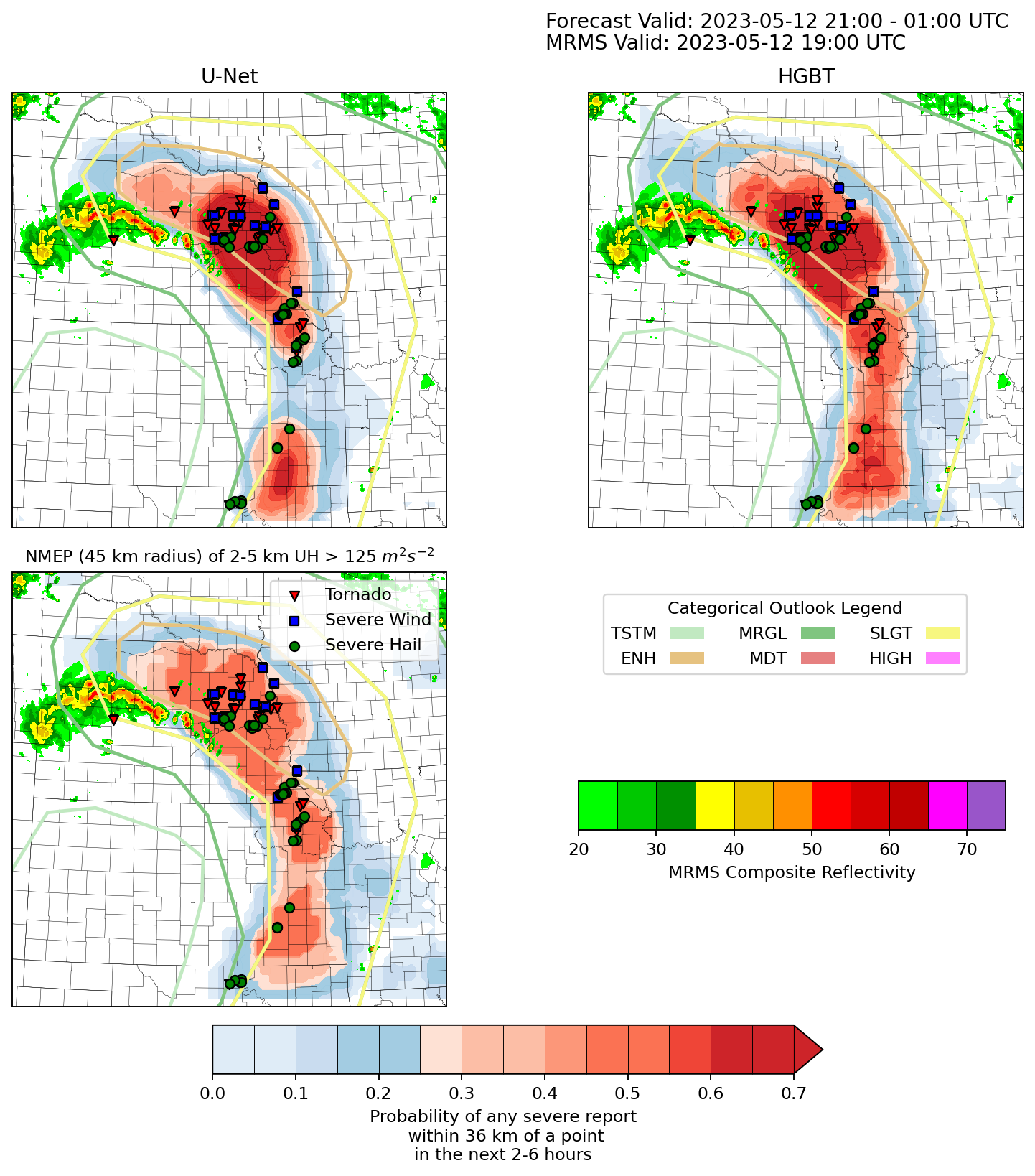}}
  \caption{Example forecasts from the U-Net (upper left), HGBT (upper right), and WoFS baseline NMEP (lower left). The guidance was issued for 12 May 2023 22-01 UTC from WoFS forecasts initialization at 20:00 UTC. SPC categorical outlooks issued at 1630 UTC the same day are overlaid as contours. The MRMS composite reflectivity at model initialization time is also shown. NCEI Storm Data reports of severe hail, severe wind, and tornadoes are shown in green, blue, and red markers, respectively. }\label{fig:example_forecasts}
\end{figure}

Figure~\ref{fig:example_forecasts} presents example forecasts from the two ML models and the WoFS baseline. This example illustrates the guidance's broad characteristics and helps ground the objective verification in the following section. In this case, SPC issued an enhanced risk focused on the Nebraska and Iowa border, while the WoFS domain was centered on the Kansas and Nebraska border. The guidance displayed is valid for 22--01 UTC on 12 May 2023. For further context, we show the MRMS composite reflectivity during forecast initialization. All three forecasts generally capture the area of severe reports in northeastern Nebraska and the isolated reports near southeastern Nebraska and southeastern Kansas, with probabilities often exceeding 40--50\%. The WoFS is a refined CAM ensemble system designed for short-term thunderstorm prediction, making it unsurprising that the baseline NMEP effectively identifies the broad severe weather areas. Although the U-net and HGBT indicate similar regions, the spatial coverage is more accurate and associated with higher probabilities (over 70\%) roughly centered on the main concentration of reports. Despite the ML guidance being more spatially precise, notable displacement errors persist. This is expected since the ML models in this study are post-processing tools and inherit the underlying prediction errors from the WoFS ensemble output. We found a few instances where the ML models dramatically adjusted probabilities spatially compared to the baseline NMEP, but this was much more the exception than the rule. The U-Net probabilities are smoother than those from the HGBT, which is unsurprising since U-Nets typically generate smoother predictions. In the U-Net predictions, there is a separation (i.e., lower probabilities) between the central corridor over Nebraska and Iowa and the elevated probabilities in southeastern Kansas, while the HGBT maintains heightened probabilities over that corridor. In other cases, however, the HGBT prediction is more discontinuous than the U-Net prediction (not shown); the U-Net seems no more likely than the HGBT to produce discontinuous forecasts that consequently miss severe weather reports.

\subsection{Performance and Reliability Diagrams}

\begin{figure}[h]
 \centerline{\includegraphics[width=35pc]{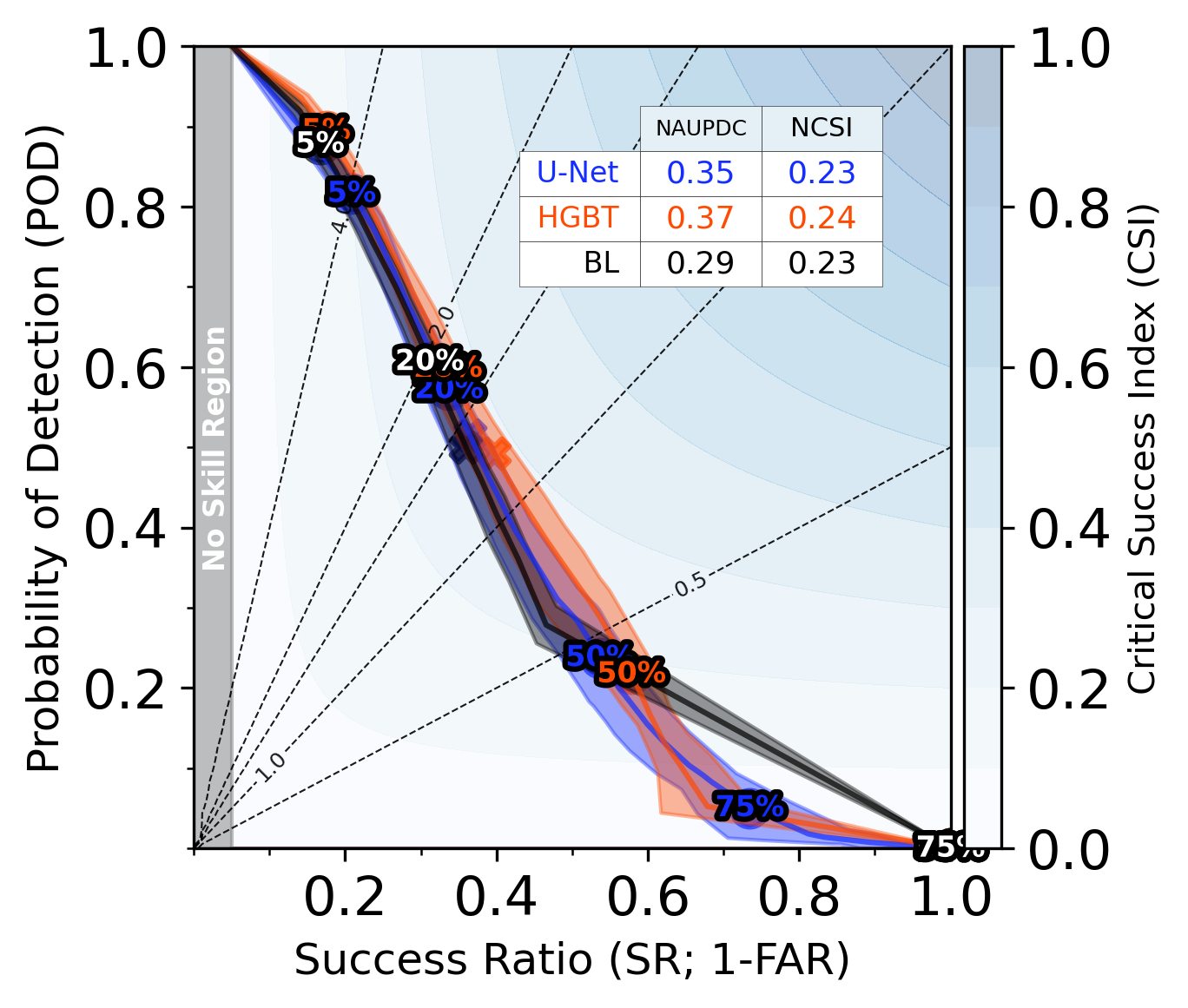}}
  \caption{The performance curves of the U-net, HGBT, and baseline (BL) are shown in blue, red, and black, respectively. POD,SR points of the three models are shown for thresholds of 5\%, 20\%, 50\%, and 75\%. The 95\% confidence interval is provided for each curve. The no-skill region, defined by the base rate, is shaded. The NAUPDC and maximum NCSI are provided in the inset table. }\label{fig:performance}
\end{figure}

\begin{figure}[h]
 \centerline{\includegraphics[width=35pc]{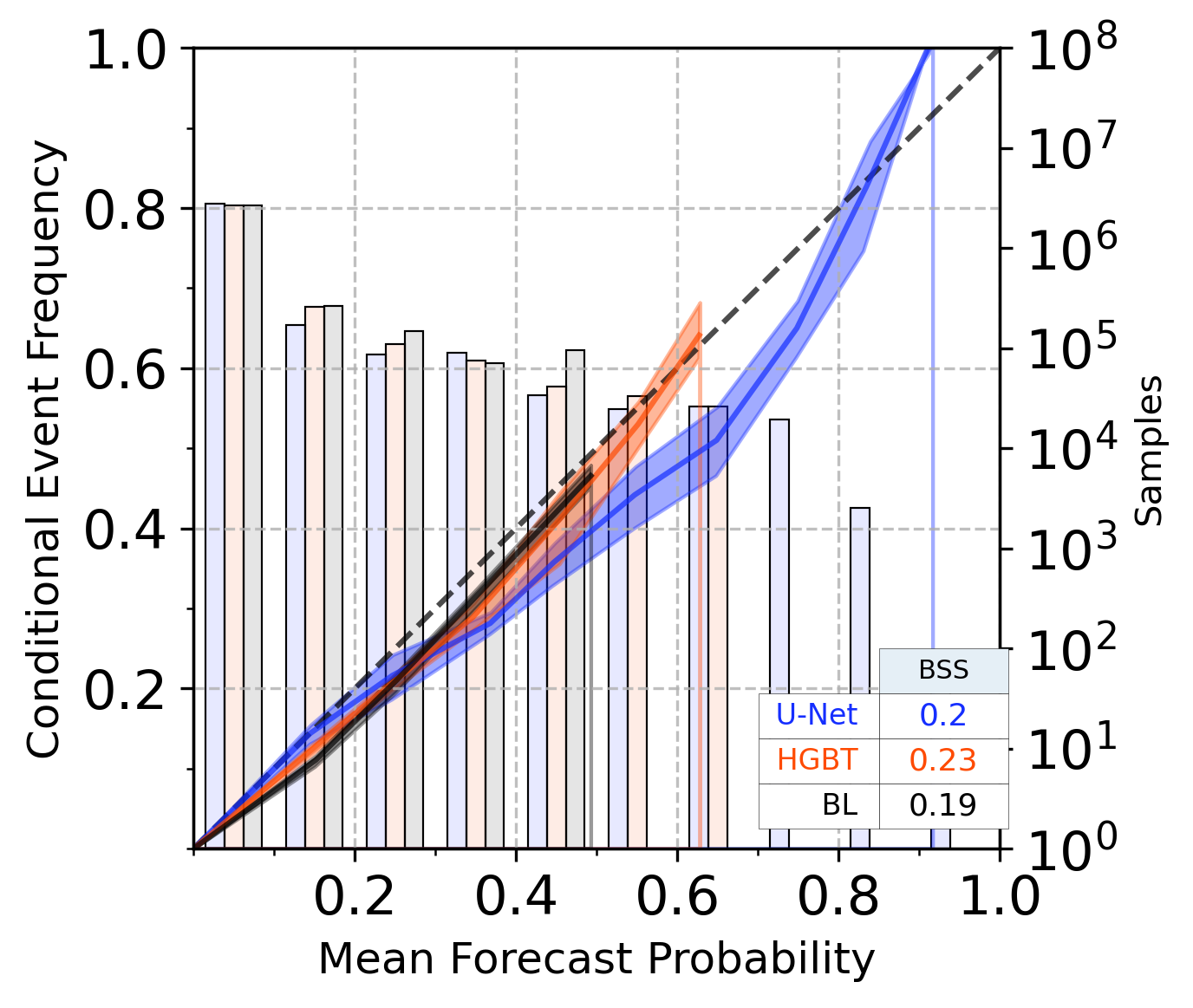}}
  \caption{The reliability curves of the U-net, HGBT, and baseline (BL) are shown in blue, red, and black, respectively. The background histograms provide the sample size in each bin (defined every 10\%) for all three models (values shown the right y-axis). }\label{fig:reliability}
\end{figure}

Figures ~\ref{fig:performance} and ~\ref{fig:reliability} show the performance and reliability diagrams. The U-Net and HGBT outperformed the baseline, yielding higher NAUPDC ($p< 0.0001$). Maximum NCSI values are comparable across the models, consistent with the previous analysis; all three models tend to highlight similar spatial areas, resulting in comparable POD and SR at a 20--25\% probability threshold. The improved performance of the ML models compared to the baseline is more evident at higher probability thresholds (40--60\%). The HGBT achieved a higher SR for those decision thresholds than the U-Net, which accounts for the slightly higher NAUPDC. Although the difference was statistically significant, whether the HGBT is meaningfully more skillful than the U-Net remains unclear.

Regarding calibration, all three models were close to the 1-to-1 line and yielded similar BSS, but both the U-Net and HGBT produced statistically significantly higher BSS ($p<0.001$). Given that all three probabilities were calibrated using isotonic regression, it is not surprising that each is generally well-calibrated. Among the three, the U-Net is the least calibrated as it tends to overpredict probabilities between 30--80\%. However, the U-Net is the only model capable of predicting 100\% probabilities, while the HGBT caps out at 60\% and the baseline at 50\%. The utility of this increased prediction range depends on the decision-making context but could be viewed favorably by end users.

%% file: d-summary.tex
\section{Summary}

Severe weather forecasting during the watch-to-warning period is challenging due to limitations of observation-based nowcasting and the inability of operational numerical guidance to explicitly predict hazards such as tornadoes, large hail, and damaging winds. A valuable forecasting tool during the watch-to-warning period is NSSL's WoFS, which provides rapidly-updating, storm-scale ensemble guidance \citep{gallo2022exploring, heinselman2024warn}. WoFS-based ML models have recently been developed to predict severe weather guidance at very short lead times (0--3 h). In the present study, we have built upon previous work to create a grid-based ML guidance system for 2--6‑h lead times. Given the success of both traditional and deep learning approaches with WoFS data \citep{Flora+etal2021, Clark+Loken2022, PotvinMLErrors2024, UnetHailTobias}, we compared a conventional tabular method (histogram gradient boosted tree [HGBT]) with a deep learning U-net framework against a carefully calibrated baseline from WoFS. 

We found that both the traditional ML and deep learning frameworks outperformed the WoFS-derived baseline. The U-net and HGBT produced statistically significantly higher NAUPDC and BSS on the held-out testing dataset. We attributed the higher skill of the ML models to their higher POD (relative to the baseline) for probabilities $>$40-50\%. Otherwise, the baseline model highlighted similar spatial areas as the ML models. This is unsurprising as the ML models are acting as post-processing tools and will inherit the underlying WoFS ensemble model errors. The ML models and the baseline were all reasonably calibrated, but the HGBT was the most calibrated. The U-Net, though the least calibrated of the three, produced probabilities up to 100\% (consistent with observed frequencies) while the HGBT was capped at 60\% and the baseline at 50\%. The utility of this increased prediction range depends on the decision-making context, but would likely be viewed favorably by some end users. Though the HGBT produced higher NAUPDC and BSS than the U-Net, whether the improvements were meaningful (e.g., would enable more accurate or longer lead-time forecasts) remains unclear. 

This study has several limitations and opens avenues for future work. First, the inputs were not identical for the U-net and HGBT models. The U-net included MRMS composite reflectivity, and the HGBT had manually defined multi-scale inputs. It is unclear whether the MRMS composite reflectivity was an important feature and how it impacted model performance. Furthermore, though the U-net is designed to capture larger-scale context through the pooling layers, it is clear how that relates to the multi-scale input given to the HGBT. Though the HGBT and U-net perform similarly, how their inputs relate to each other is questionable. Ultimately, our goal was not to determine whether HGBT or U-net was the superior approach for our prediction task. A more comprehensive comparison of the performance of deep learning versus traditional ML for WoFS-based severe weather prediction exceeds the scope of this study. It would be necessary to analyze multiple datasets and attempt to optimize the deep learning model further to conclude the advantages and disadvantages of each method.  Second, the training and verification were based on local storm reports, which introduces known limitations.  Future work should explore whether these biases manifest in the predictions, similar to \citet{Flora+etal2025}. Third, we did not explore model explainability and whether the U-net or HGBT were learning similar relationships from the data. A detailed investigation of explainability is beyond the scope of this study. Fourth, although this study used a range of ensemble statistics for the input features, little research has explored which statistics extract the most useful information from the ensemble. Future work should explore how to properly leverage ensemble information when post-processing CAM ensemble output.  Lastly, as discussed for the U-net calibration in Section~\ref{sec:ml_methods}\ref{sec:unet}, end-user feedback is required for further model development. Future work could explore whether the sharpness or higher confidence of the U-net enhances its value to forecasters.

%% file: e-appendix.tex
\appendix
\appendixtitle{Model Training Details}

\textbf{Feature Normalization.}
Input features for the histogram gradient-boosted tree (HGBT) and the U-Net were normalized using the mean and standard deviation computed from the training dataset. Separate normalization statistics were calculated for the tabular HGBT dataset and the DL input grids to ensure consistency within each framework.

\textbf{HGBT Hyperparameter Optimization.}
 HGBT model hyperparameters were optimized using Bayesian hyperparameter optimization via the hyperopt Python package \citep{Bergstra2013}, with five-fold cross-validation. Optimization was performed using AUPDC (see Section~\ref{sec:ml_methods}\ref{sec:metrics}) as the scoring metric. To improve efficiency, we implemented an early stopping condition requiring a 1\% performance improvement within every 10 iterations. Hyperparameter ranges and selected values (bold) are provided in Table ~\ref{table:hyperopt}. The default values from scikit-learn v1.0.2 \citep{pedregosa2011scikit} were used for unlisted parameters.

\textbf{Deep Learning Hyperparameters.}
The U-Net model hyperparameters were selected using a random search ($N$=100). This sampling-based approach has shown success in previous DL studies (e.g., \citealt{Chase+etal2024}). The hyperparameter ranges are provided in Table ~\ref{table:hyperopt}. The ReLU activation function was used. The initial filter size is doubled for subsequent layers; for example, given an initial filter size of 64 and a 3-layer U-net, the filters per layer are 64, 128, and 256, respectively. For a given iteration, the maximum epoch was 200, but we used early stopping based on validation loss with a patience of 7. The best iteration trained for 130 epochs, however, the model checkpoint used for the final evaluation was the iteration with the highest AUPDC on the validation dataset. 
The initial learning rate was prescribed, but we used a decaying learning rate scheduler based on validation performance (ReduceLROnPlateau). To accelerate model training, bfloat16 precision was used. The final model size is 32M parameters.

\begin{table}[t]
\caption{Hyperparameter values attempted for the HGBT in the hyperparameter search. Selected values are bolded.} 
\label{table:hyperopt}

\begin{center}
\begin{tabular}{ll}
\hline \hline

Hyperparameter & Values   \\ 

\hline

Learning Rate & 0.0001, 0.001, 0.01,\textbf{ 0.1} \\
Max. Leaf Nodes & \textbf{5}, 10, 20, 30, 40, 50 \\
Max. Depth &  4, \textbf{6}, 8, 10 \\
Min. Samples Leaf & 5,\textbf{10},15,20,30, 40, 50 \\
L2 Regularization & 0.001, \textbf{0.01}, 0.1 \\
Max. Bins & 15, 31, \textbf{63}, 127 \\

\hline
\end{tabular}
\end{center}
\end{table}


\begin{table}[t]
\caption{Hyperparameter values attempted for the U-net in the hyperparameter search. Selected values are bolded.} 
\label{table:dl_hyperopt}

\begin{center}
\begin{tabular}{ll}
\hline \hline

Hyperparameter & Values   \\ 

\hline

Kernel Size  & 3 \\ 
Dropout Rate & $\mathcal{U}(0, 0.4)$ $\rightarrow$ \textbf{0.3} \\
Initial Filter Size & 32, \textbf{64}, 128, 256 \\
Num. of Layers & 1,2,3,\textbf{4} \\
Learning Rate & 0.0001, \textbf{0.001}, 0.01, 0.1 \\

\hline
\end{tabular}
\end{center}
\end{table}

\clearpage